\documentclass[12pt,fleqn]{article}
\usepackage{psfig}
\begin{document}
\begin{center}
{\Large{\bf PHOTOPRODUCTION OF THE $\Lambda(1405)$ ON THE PROTON AND NUCLEI}}
\end{center}

\vspace{1cm}

\begin{center}
{\large{ J.C. Nacher$^ {1,2}$, E. Oset$^ {1,2}$, H. Toki$^1$ and A. Ramos$^3$}}
\end{center}

\vspace{0.4cm}
%{\it $^1$  Departamento de F\'{\i}sica Te\'orica and IFIC 

%Centro Mixto Universidad de Valencia-CSIC
%46100 Burjassot (Valencia), Spain.}       
\begin{center}
{\it $^1$ Research Center for Nuclear Physics (RCNP), Osaka University,}
%\end{center}
%\begin{center}
{\it Ibaraki, Osaka 567-0047, Japan.}
\end{center}
\begin{center}
{\it $^2$  Departamento de F\'{\i}sica Te\'orica and IFIC, 
Centro Mixto Universidad de Valencia-CSIC
46100 Burjassot (Valencia), Spain.}
\end{center}
\begin{center}
{\it $^3$ Departament d'Estructura i Constituents de la Materia, Universitat de
Barcelona,}
%\end{center}
%\begin{center}
{\it Diagonal 647, 08028 Barcelona, Spain.}
\end{center}
\vspace{2.2cm}

\begin{abstract}
{\small{We study the 
$\gamma p\rightarrow  K^+ \Lambda(1405)$ reaction at energies close to
threshold using a chiral unitary model where the resonance is
generated dynamically from $K^-p$ interaction with other channels constructed
from the octets of baryons and mesons.
 Predictions are made for cross sections into several channels and it is
 shown that the detection of the $K^+$ is sufficient to determine the shape
 and strength of the $\Lambda(1405)$ resonance. The determination of the 
 resonance properties in nuclei requires instead the detection of the resonance
 decay channels. Pauli blocking effects on the resonance, which have been shown to be
 very important for the resonance at rest in the nucleus, are irrelevant here where
 the resonance is produced with a large momentum. The nuclear modifications here
 would thus offer information on the resonance and $K^-$ nucleus dynamics complementary
 to the one offered so far by $K^-$ atoms.}} 
\end{abstract}
\vspace{2.2cm}
\newpage
The $S_{01}$ $\Lambda$(1405), I($J^P$)=0($\frac{1}{2}^-$), resonance is situated just below the
$K^-N$ threshold and plays an important 
role in the $\bar{K}N$ interaction at low
energies [1]. The presence of this resonance has an important role in the
properties of kaons in nuclear matter and nuclei. Modifications of the
resonance properties in nuclei lead to substantial changes in the kaon
selfenergy in the medium and have repercussions in  atoms which have been
   often emphasized [2, 3]. 
   
   The $\Lambda$(1405) resonance is well described within a
   unitary coupled channel approach, as shown earlier in [4, 5] and more recently
   in [6, 7, 8, 9]. An important step in the understanding of the $\Lambda$(1405)
   dynamics has been done in [7, 8, 9] where, using  chiral Lagrangians 
   and a unitary coupled channel approach, the properties of the resonance , as
   well as the low energy $K^-N$ cross sections, are well reproduced. The use of
   nonperturbative unitary approaches in coupled channels shows here its strength
   over ordinary perturbation schemes which would be unable to generate a resonance
   structure in the scattering matrix.

   In ref. [9], where the whole set of mesons of the pseudoscalar octet,
   as well as those of the octet of stable baryons, are used, one can reproduce
   all the low energy $K^-p$ data, as well as the properties of the
   $\Lambda(1405)$ resonance
    by means of the lowest order chiral Lagrangian with only one
   parameter, a suitable cut off. This cut off is needed to regularize the loop integrals in the
   Bethe Salpeter equation and, at the same time, allows to effectively 
   incorporate effects of the higher order chiral Lagrangians. 
    These findings, similar to those in [7, 8], favour the interpretation
    of the latter resonance as a quasibound state of mesons and baryons, different
    in nature to genuine baryons of 3q internal structure which would survive
    in the large $N_C$ limit [10].

     One of the interesting findings in the  nuclear modifications of the $\Lambda$(1405)
   resonance is the effect of Pauli blocking in the intermediate states of the
   coupled channels, which moves the resonance to higher energies and changes the
   sign of the $K^-p$ amplitude at threshold from repulsive to attractive [11,
   12].
   The attraction in the medium is suggested  by the analysis of the the data 
   of kaonic atoms [2, 3, 13, 14]. This attraction in the medium could eventually lower
   the kaon mass sufficiently to produce kaon condensates in a neutron star
   [15]. However, a recent selfconsistent determination of the $K^-$ selfenergy in
   the medium, accounting for Pauli blocking and the effect of the kaon
   selfenergy itself, leads to substantially different results in the sense that
   the position of the resonance barely moves in the medium [16]. It becomes 
   wider while some attraction on the kaons is also generated.
   A more recent calculation, incorporating in addition the selfenergy of the
   pions in the intermediate $\Sigma \pi$ and $\Lambda \pi$ channels, comes to support
   these latter findings about the position, only the width becomes larger and
   at densities close to nuclear matter the resonance structure disappears
   [17]. A 
   net attraction on the $K^-$ is generated in this approach at densities higher than 
   $\rho_0/10$.

     As one can see, the theoretical situation is at a stage where experimental
   information is needed to settle the questions, which are also essential to
   put the hypothesis of the kaon condensation on firmer grounds.
     The present work suggests a nuclear reaction where an answer to some of
   these questions can be obtained. The reaction is photoproduction of the
   $\Lambda(1405)$ resonance on the proton and in nuclei, which can be easily implemented
   in present facilities like TJNAF and LEPS of SPring8/RCNP.  
     The elementary reaction is 
%\newline
\begin{equation}
%\hspace{2cm}
\gamma p\rightarrow  K^+ \Lambda(1405) .
\end{equation}     
%\newline
\hspace{0.5cm}
      In order to get the coupling of the mesons and
   the photon to the $\Lambda$(1405) resonance we make use of the model of [9] 
   for meson-baryon interaction which
   generates the resonance dynamically from the 10 coupled channels, $K^-p$,
    $\bar{K}^0 n$, $\pi^0\Lambda$, $\pi^0 \Sigma^0$, $\eta \Lambda$,
     $\eta \Sigma^0$, $\pi^+ \Sigma^-$, $\pi^- \Sigma^+$, 
   $K^+ \Xi^-$, $K^0 \Xi^0$. The starting
   point is the meson-baryon amplitude originated from the
   lowest order chiral Lagrangian [18, 19, 20] which generates the coupling for
   the $M_iB_i \rightarrow M_jB_j$ transition [9]
\begin{equation}
V_{ij} = - C_{ij}\frac{1}{4f^2}\overline{u}(p')\gamma^\mu u(p)(k_\mu + k'_\mu)
,
\end{equation} 
where $f$ is the decay constant of the pion ($f_\pi$ = 93 MeV). In [9] this constant is
   taken as an average between the one for pions and kaons, $f$ = 1.15$f_\pi$. In eq. (2)
   $k_\mu$ and $k'_\mu$ are the initial and final meson momenta and $C_{ij}$ is a
   10$\times$10
   matrix which is shown in table 1 of [9].
     For the low energies which we will consider here only the s-wave is
   relevant and the expression of eq. (2) is further simplified to 
\begin{equation}
V_{ij} = - C_{ij}\frac{1}{4f^2}(k^0 + k'^0) .
\end{equation}

     The scattering matrix between two channels, $T_{ij}$, is then given by means
   of the Bethe Salpeter equation, where V and T are shown to factorize on
   shell outside the loop integral using the heavy baryon approximation [9]. Hence one has the algebraic matrix
   equation
\begin{equation}
T = V + VGT \rightarrow T = [1 - VG]^{-1}V   
\end{equation}
  {
   with G a diagonal matrix, with matrix elements
    
\begin{equation}  
G_{l}  =  i\int\frac{d^4q}{(2\pi)^4}\, \frac{M_l}{E_l(\vec{q})}
\, \hspace{0.2cm}\frac{1}{k^0 + p^0 - q^0 - 
E_l(\vec{q}) + i\epsilon}\, \hspace{0.2cm} \frac{1}{q^2 - m_l^2 + i\epsilon}
\end{equation}
corresponding to a loop diagram with an intermediate meson and baryon
   propagators and which is regularized with a cut off in $\vec{q}$, with
   $q_{max}$ = 630 MeV in [9], the only parameter fitted to the data.

      The $\Lambda(1405)$ is a s-wave resonance and thus we concentrate on
      mechanisms which allow s-wave baryon-meson production.	     
      In a first step the reaction would proceed as depicted diagrammatically
      in Fig. 1 which forms a gauge invariant set of diagrams at threshold
      of the pair meson production. In order to simplify the
theoretical calculations we choose a photon energy around 
$E_{\gamma Lab}$ = 1.7
   GeV where the particles are produced with low energy. In that case it is
   easy to prove that the contribution of the meson pole terms (diagrams a, b,
   of Fig. 1) is negligible (less than 5$\%$ contribution) and one can live with
   the contact term alone (diagram c). This choice of energy is also suited
   theoretically to prevent the contribution of terms like in Fig. 1a, where
   the photon couples directly to the nucleon (or excites resonances) and then
   one has the contact vertex MMBB. In such a case one gets the combination
   $(k'_0-k_0)$ in eq. (3) which is a very small number. Similarly, other terms
   where the mesons would be produced at different vertices would involve
   p-wave couplings which would also vanish at threshold. The smallness of the
   momenta of the mesons involved is also suited to prevent the formation  of
   a p-wave resonance like the $\Sigma$(1385). In any case the small branching ratio
   (12\%) of this resonance into the $\pi\Sigma$ channel makes easy the separation
   from the $\Lambda(1405)$ which decays only into that channel.
   
\vspace{1.5cm}
\begin{figure}[h]
\centerline{\protect
\hbox{
\psfig{file=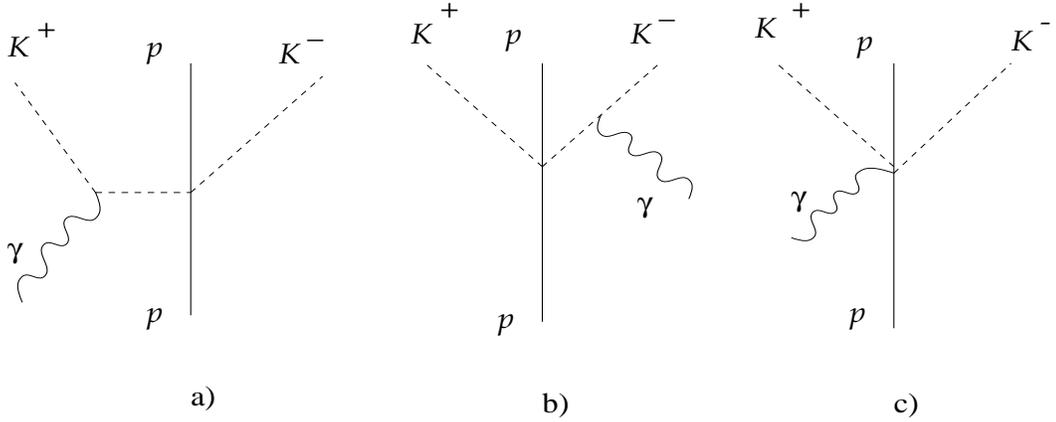,height=5.5cm,width=14.0cm,angle=-90}}}
\caption{Feynman diagrams used in the model for the $\gamma
p\rightarrow K^+\Lambda(1405)$ reaction. }
\label{Fig.1}
\end{figure}
     Minimal coupling from eq. (2) provides the contact term $\gamma M_iB_i 
\rightarrow M_jB_j$ 

\begin{equation}
V_{ij}^{(\gamma)} = C_{ij}\frac{e}{4f^2}(Q_i + Q_j)\, 
\overline{u}(p')\gamma^\mu u(p)\epsilon_\mu ,
\end{equation} 
where $Q_i$, $Q_j$ are the initial and final meson
charges and $\epsilon_\mu$ the photon polarization vector.

     In this situation close to threshold which we have chosen we can neglect terms proportional to the final baryon momentum,
     which are small in the kinematics chosen. In this case the contact term of Fig. 1c
     for $\gamma p \rightarrow K^+ M_j B_j$ (or equivalently $\gamma p K^-\rightarrow M_j
     B_j$ ), where $M_j B_j$ is any of the 10 states that couple to the
     $\Lambda$(1405), 
     can be written in the Coulomb gauge $\epsilon^0=0$, $\vec{\epsilon}\cdot\vec{q}=0$  
      and in the $\gamma p$ CM frame as
%      the contact vertex of Fig. 1c,
%     where in the final state we have a $K^+$ and any of the 10 meson-baryon
%     states that couple to the $\Lambda$(1405) and which we have described at the beginning,
%   the contact term in the CM of the $\gamma$-p system , and in the Coulomb gauge
%   $\epsilon_0=0$, $\vec{\epsilon}\cdot\vec{q}=0$, can be written for the 
%   $\gamma p\rightarrow K^+ M_j B_j$ reaction as

\begin{equation}
V_{j}^{(\gamma)} = C_{1j}\frac{e}{4f^2}
\, \frac{i(\vec{\sigma}\times\vec{q})\vec{\epsilon}}{2M}\, Q'_j ,
\end{equation}
where the index 1 in $C_{1j}$ stands for the $K^- p$ channel and 
   $Q'_j$ =$1 - Q_j$. Here 
   $Q'_j$ is the vector (2, 1, 1, 1, 1, 1, 0, 2, 0, 1) 
   corresponding to the final
   states in the order mentioned in the introduction and M stands for the
   proton mass.
   
    The $\Lambda$(1405) resonance is generated dynamically by iteration of the 
    MB$\rightarrow$ M'B'
    vertex according to the Bethe Salpeter equation, and diagrammatically
    this is shown in Fig. 2.

\begin{figure}[h]
\centerline{\protect
\hbox{
\psfig{file=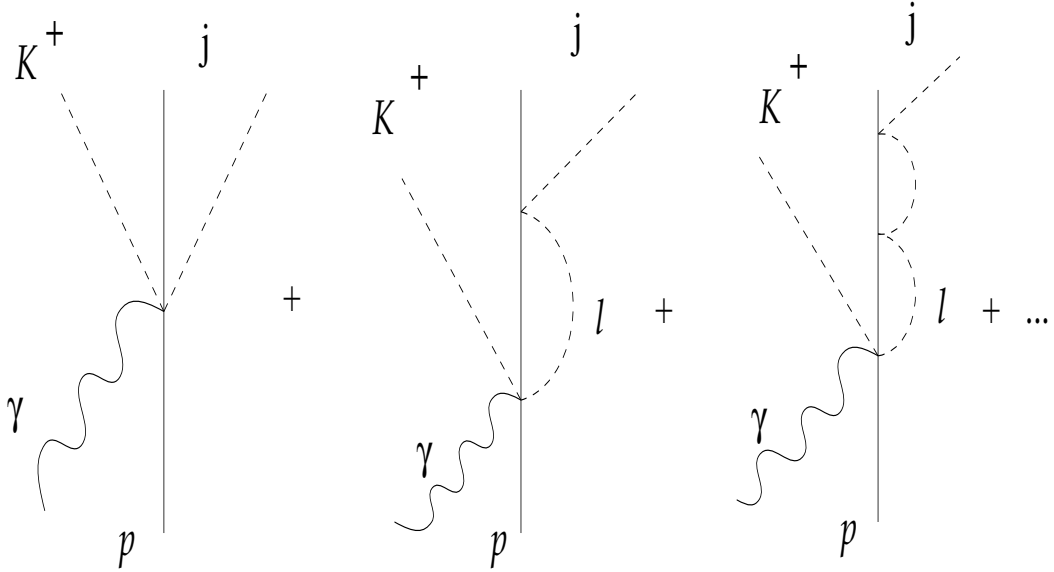,height=7.5cm,width=14.0cm,angle=-90}}}
\caption{Diagrammatic representation of the meson-baryon final state interaction in the
 $\gamma p\rightarrow
K^+\Lambda$(1405) process. }
\label{Fig.1}
\end{figure}

     The sum implicit in Fig. 2 is easily evaluated. The $t^{(\gamma)}_j$ matrix for
     the process with the j channel in the final state is given by means of
     the vector $D_j = C_{1j}\cdot Q'_j$ as:
\begin{equation}     
t^{(\gamma)}_{j} = 
\frac{e}{4f^2}\, \frac{i(\vec{\sigma}\times\vec{q})\vec{\epsilon}}{2M}\, (D_j
+ \sum_l D_l G_l T_{lj}) ,
\end{equation} 
where the on shell factorization of the strong amplitude found in [9] and in
[21] for the related electromagnetic process has been used.
    The particular structure of eq. (8) allows one to obtain an easy formula
    for the invariant mass distribution of the final j state, particularly
    suited to the search of a resonance. We find

\begin{equation} 
%\newline
\hspace{-1cm}
\frac{d\sigma}{dM_I}\left.\right|_j = \frac{1}{(2\pi)^3}\frac{1}{4s}\frac{MM_j}
{s-M^2}\frac{1}{M_I}
\overline{\sum}\sum|t^{(\gamma)}_j|^2\lambda^{1/2}(s, M_I^2, m_K^2)\lambda^{1/2}(M_I^2, M_j^2,
m_j^2) 
\end{equation}
where $M_I$ is the invariant mass of the j state , $M_j$, $m_j$ 
   the masses of the 
   j state and $\lambda$(x, y, z) the ordinary K\"{a}llen function.
   
    In Fig. 3 we show $\frac{d\sigma}{dM_I}$ for the different channels. 
While all coupled channels collaborate to the built up of the 
$\Lambda$(1405) resonance, most of them open up at higher energies and the resonance
shape is only visible in the
 $\pi^+\Sigma^-$,
     $\pi^-\Sigma^+$,
   $\pi^0\Sigma^0$ channels. The  $\bar{K}N$ production occurs at
   energies slightly above the resonace and the $\pi^0 \Lambda$, with isospin one, only
   provides a small background below the resonance.

     It is interesting to see the different shapes of the three $\pi\Sigma$ 
     channels.
     This can be understood in terms of the isospin decomposition of the states
\begin{equation}
    |\pi^+\Sigma^-\rangle = - \frac{1}{\sqrt{6}}|2, 0\rangle - \frac{1}{\sqrt{2}}|1, 0\rangle -
     \frac{1}{\sqrt{3}}|0, 0\rangle
\end{equation}
\begin{equation}
    |\pi^-\Sigma^+\rangle = - \frac{1}{\sqrt{6}}|2, 0\rangle + \frac{1}{\sqrt{2}}|1, 0\rangle -
     \frac{1}{\sqrt{3}}|0, 0\rangle
\end{equation}
\begin{equation}
|\pi^0\Sigma^0\rangle =  \sqrt{\frac{2}{3}}|2, 0\rangle - \frac{1}{\sqrt{3}}|0, 0\rangle 
\end{equation}

   Disregarding the I = 2 contribution which is negligible, the cross sections
   for the three channels go as:
\begin{equation}
\frac{1}{2}|T^{(1)}|^2 + \frac{1}{3}|T^{(0)}|^2 
+ \frac{2}{\sqrt{6}}Re(T^{(0)} T^{(1)*})\, ;\hspace{0.5cm} \pi^+\Sigma^-
\end{equation}
\begin{equation}
\frac{1}{2}|T^{(1)}|^2 + \frac{1}{3}|T^{(0)}|^2 
- \frac{2}{\sqrt{6}}Re(T^{(0)} T^{(1)*})\, ;\hspace{0.5cm} \pi^-\Sigma^+
\end{equation}
\begin{equation}
\frac{1}{3}|T^{(0)}|^2\, ;\hspace{0.5cm} \pi^0\Sigma^0
\end{equation}

    The crossed term $T^{(0)}T^{(1)*}$ is what makes these cross sections different. We can
    also see that

\vspace{0.5cm}
\hspace{0.5cm}
 3$\frac{d\sigma}{dM_I}(\pi^0\Sigma^0)\simeq \frac{d\sigma}{dM_I}(I=0)$
\\

\begin{equation}
\frac{d\sigma}{dM_I}(\pi^0\Sigma^0) + \frac{d\sigma}{dM_I}(\pi^+\Sigma^-) +
\frac{d\sigma}{dM_I}(\pi^-\Sigma^+)\simeq \frac{d\sigma}{dM_I}(I=0) 
+  \frac{d\sigma}{dM_I}(I=1)
\end{equation}

   This means that the real shape of the resonance must be seen in either the
   $\pi^0 \Sigma^0$ channel or in the sum of the three $\pi\Sigma$ channels,  provided the I
   = 1
   cross section (not the crossed terms which are relatively large ) is small
   as it is the case. Incidentally, eqs. (13, 14) also show that the difference between the
   $\pi^+ \Sigma^-$ and $\pi^- \Sigma^+$ cross sections gives the crossed term and hence
   provides some information on the I = 1 amplitude.

	In Fig. 4 we recombine the results in a practical way from the experimental point of
	view. We
	 show the I = 0 contribution, the $\Sigma^0 \pi^0$ contribution and the sum of all channels including
	the $\pi^0 \Lambda$, and we see that they are all very similar and the total contribution is just the
	$\Lambda$(1405) contribution plus a small background. In practical terms this result means
	that the detection of
	the $K^+$ alone (which sums the contribution of all channels) is sufficient to determine the shape and the
	strength of the $\Lambda$(1405) resonance in this reaction.

	The study of the reaction in nuclei requires special care. Indeed, assume one uses the same
	set up as before with a nuclear target and measures the outgoing $K^+$. There the invariant
	mass will be given by
\begin{eqnarray}	
 M_I\, ^2(p) = (q + p - k)^2 & = & M^2 + m_K^2 - 2 q^0 k^0 +
  2\vec{q}\cdot\vec{k} + \\ \nonumber
& & 2 p^0 (q^0 - k^0) -
2 \vec{p}\cdot(\vec{q} - \vec{k})
\end{eqnarray}  
with $q, k, p$ the momenta of the photon, $K^+$ and initial proton respectively.
	Since $\vec{q}-\vec{k}$ has a large size, there will be a large spreading of invariant
	masses due to Fermi motion for a given set up of photon and $K^+$ momenta, unlike
	in the free proton case where $M_I\, ^2$ is well determined\footnote{
	We are endebted to T. Nakano and J. K. Ahn for calling us the attention on this point}. The nuclear cross section
	normalized to the number of protons (the neutrons through $K^-n$ and coupled channels
	only contribute to I = 1 with a small background) would be given by the convolution 
	formula

\begin{equation}
\frac{1}{Z}\frac{d\sigma}{dM_I}\bigm|_A\simeq\frac{2}{\rho_p}\int\frac{d^3p}{(2\pi)^3} \,
\frac{d\sigma}{dM_I(\vec{p})}\hspace{.6cm} ; \hspace{1.cm} \rho_p =\frac{k_F\, ^3}{3\pi^2}
\end{equation}	
where the integral over $\vec{p}$ ranges up to the Fermi momentum $k_F$.
	
	In order to show the effects of the Fermi motion we choose values of $\vec{k}$ 
	corresponding to forward $K^+$ in the CM (and hence largest value of $\vec{k}$ 
	in the lab frame) which would minimize the spreading of the $M_I\, ^2(\vec{p})$ in eq. (17).
	Even then, we can see in Fig. 4 the result of the distribution. The $M_I$ in the
	x axis of the figure in this case is taken for reference from eq. (17) for a nucleon at
	rest. We can see that the spreading of the invariant masses is so large that one looses
	any trace of the original resonance. This simply means that in order to see genuine
	dynamical effects one would have to look at the invariant mass of the resonance
	from its decay product, $\pi\Sigma$, tracing back this original invariant mass with appropiate
	final state interaction corrections.

		One interesting thing here is that the $\Lambda(1405)$ resonance is produced
		with a large momentum in the nuclear lab frame. We have checked that Pauli
		blocking effects in the resonance, which were so important for the resonance
		at rest in the nucleus, become now irrelevant. Hence medium modifications
		of the resonance in the present situation should be attributed to other
		dynamical effects [17].
		
		The flattering of the resonance in nuclei when only the $K^+$ is detected
		has its positive aspects though. Indeed by using
		$C H_2$ targets instead of just $H$, as planned in SPring8/RCNP, and detecting
		only the $K^+$, one would be observing the clear signal of the $H$ on top
		of a flat background of the $^{12} C$.

	The cross sections obtained, of the order of 5$\mu$b/GeV, are in the easily attainable range in present
	experimental
	facilities like TJNAF and LEPS of SPring8/RCNP. The results obtained in this work should encourage the
	performance of the actual experiments. They would serve to test
	current ideas on chiral symmetry from the elementary reaction and
	with nuclear targets would
	 give us much needed information on the in medium properties of the
	$\Lambda$(1405) resonance and $K^-$. This should help
	resolve questions like the $K^-$ condensation, and the origin of the attraction seen in 
        $K^-$ atoms.

\vspace{3cm}

{\bf Acknowledgements.}
We acknowledge useful discussions and comments from A. Titov, T. Nakano and J. K.
Ahn. We are grateful to the COE Professorship program of Monbusho, which enabled E. O. to
stay at RCNP to perform the present work. One of us, J.C. Nacher would like to acknowledge the hospitality of the RCNP of the Osaka University where
	this work was done and support from the Ministerio de Educacion y Cultura. This work is partly supported
	by DGICYT contract number PB96-0753 and PB95-1249.

\newpage
{\Large{\bf Figure captions}}
\\

Figure 3: Mass distribution for the different channels. The $\Sigma^+\pi^-$,
$\Sigma^-\pi^+$ and $\Sigma^0\pi^0$ distributions are shown in the figure with
the dashed lines. The solid line with the resonance shape is the sum of the
three $\Sigma\pi$ channels divided by three. In addition the distributions for
$\pi^0\Lambda$ and $K^-p$ production are also shown. The $\bar{K^0}n$
production is small and not shown there.
\\

Figure 4: Mass distributions. Dashed line (resonant
shape): $\Sigma^0\pi^0$
 distribution multiplied by three. Dotted line (resonant shape): pure I = 0 contribution from
 the $\Sigma\pi$ channels. Solid line (resonant shape): Sum of the cross sections for all the
 channels. In addition the I = 1 background contribution from the $\Sigma\pi$
 and $\pi^0\Lambda$ channels is also shown. Short-dash-dotted line: Effects of
 the Fermi motion with $k_F = 268 MeV/c$ $(\rho = \rho_0)$ where the free space
 $\Lambda(1405)$ distribution is assumed in the calculation.
\\

\end{document}